\documentclass[12pt]{article}
\usepackage{amsmath,amsfonts,amssymb,amsthm,amstext,amscd,eucal}
\usepackage[all]{xy}

\makeatletter \@addtoreset{equation}{section}

\makeatletter\renewcommand\section{\@startsection {section}{1}{\z@}%
                                   {-3.5ex \@plus -1ex \@minus -.2ex}
                                   {2.3ex \@plus.2ex}%
                                   {\normalfont\large\bfseries}}
\renewcommand\subsection{\@startsection{subsection}{2}{\z@}%
                                     {-3.25ex\@plus -1ex \@minus -.2ex}%
                                     {1.5ex \@plus .2ex}%
                                     {\normalfont\bfseries}}

\usepackage{amsmath,amsfonts,amssymb,amsthm,amstext,bbm, amscd,eucal}
\usepackage{epsfig}
\usepackage{graphicx}
\usepackage{psfrag}
\usepackage{epstopdf}

\newcommand{\ie}{{\em i.e. }}

\newcommand{\be}{\begin{equation}}
\newcommand{\ee}{\end{equation}}
\newcommand{\bea}{\begin{eqnarray}}
\newcommand{\eea}{\end{eqnarray}}

\newcommand{\bse}{\begin{subequations}}
\newcommand{\ese}{\end{subequations}}
\newcommand{\bi}{\begin{itemize}}
\newcommand{\ei}{\end{itemize}}

\begin{document}
\begin{titlepage}

\begin{flushright}\vspace{-3cm}
{\small
\today }\end{flushright}
\vspace{-.5cm}

\begin{center}
\centerline{{\Large{\bf{Exit from Inflation with a First-Order Phase Transition}}}}  \vspace{4mm}
\centerline{{\Large{\bf{and a Gravitational Wave Blast}}}} \vspace{4mm}
{\large{{\bf Amjad~Ashoorioon\footnote{e-mail: amjad.ashoorioon@physics.uu.se}$^{,a}$}}}


\bigskip\medskip
\begin{center}
{$^a$ \it Institutionen f\"{o}r fysik och astronomi
Uppsala Universitet, Box 803, SE-751 08 Uppsala, Sweden }\\
\end{center}
\end{center}
\setcounter{footnote}{0}

\date{\today}

\begin{abstract}
In double-field inflation, which exploits two scalar fields, one of the fields rolls slowly during inflation whereas the other field is trapped in a meta-stable vacuum. The nucleation rate from the false vacuum to the true one becomes substantial enough that triggers a first order phase transition and ends inflation. We revisit the question of first order phase transition in an ``extended'' model of hybrid inflation, realizing the double-field inflationary scenario, and correctly identify the parameter space that leads to a first order phase transition at the end of inflation. We compute the gravitational wave profile which is generated during this first order phase transition. Assuming instant reheating, the peak frequency falls in the $1$ GHz to $10$ GHz frequency band and the amplitude varies in the range $10^{-11}\lesssim \Omega_{\rm GW} h^2 \lesssim 10^{-8}$, depending on the value of the cosmological constant in the false vacuum.  For  a narrow band of vacuum energies, the first order phase transition can happen after the end of inflation via the violation of slow-roll, with a peak frequency that varies from $1$ THz to  $100$ THz. For smaller values of cosmological constant, even though inflation can end via slow-roll violation, the universe gets trapped in a false vacuum whose energy drives a second phase of eternal inflation.  This range of vacuum energies do not lead to viable inflationary models, unless the value  of the cosmological constant is compatible with the observed value, $M\sim 10^{-3}$ eV. \end{abstract}

\end{titlepage}
\renewcommand{\baselinestretch}{1.1}

\section{Introduction}

The mechanism that ends inflation  is still an open question in cosmology. In fact it was the achilles heel of the old inflation model \cite{Guth:1980zm}, as the universe would never recover from inflation. Accelerated expansion of the inflationary false vacuum background  pushes the walls of true vacuum bubbles exponentially apart such that they never manage to coalesce and end inflation. New inflation \cite{Linde:1981mu}, with the evolution of the inflaton, allowed for another mechanism of termination of inflation. Fast-roll evolution of the scalar field at the end of inflation can violate the slow-roll condition and end inflation. However there still remained to address how the universe has been reheated from the supercooled phase. For that one has two assume that the inflaton is coupled to other degrees of freedom and decay of the inflaton to these degrees of freedom happens either perturbatively or nonperturbatively \cite{Kofman:1997yn}. For this to happen one often has to fine-tune the couplings or the bare-masses of these new degrees of freedom.

End of inflation through bubble nucleation has this advantage that reheating the universe happens from the collision of true vacuum bubble walls and their natural conversion to radiation. The idea of new and old inflations were combined in ``double-field'' inflation \cite{Linde:1990gz,Adams:1990ds} in which one of the fields rolls during inflation, as in slow-roll inflation, and the second field is initially trapped in the meta-stable vacuum (as can be seen in our example, this vacuum could be the only existing minimum in the beginning of inflation. The true vacuum could develop as inflation proceeds). As the first field rolls, the nucleation rate from the meta-stable vacuum to the true one becomes large enough that bubbles of true vacuum can indeed percolate and end inflation. This is very similar to the false-vacuum dominated Hybrid inflation \cite{Linde:1993cn,Copeland:1994vg} where the tachyonic instability of the waterfall field is replaced with a first order phase transition.  A prototype of such a potential takes the form
\be\label{pot-tot}
V(\phi,\psi)=V_0+V_1(\phi)+ V_2(\phi,\psi),
\ee
where $V_0$ is the vacuum energy which is constant and $\phi$ is the rolling field whose potential $V_1(\phi)$ along with the vacuum energy drives inflation. $\psi$ is the field which facilitates the first order phase transition to the true vacuum.

In this paper we focus on an example of such double-field inflationary models, realized in the context of extended hybrid inflation  \cite{Copeland:1994vg}.  Previously, \cite{Cortes:2009ej} had  tried to identify the parameter space which allows for exit from inflation with a first order phase transition. In their calculation, they had assumed that the pre-exponential factor  in the nucleation rate per unit four-volume, $\Gamma=\mathcal{A} \exp(-S_{E})$, is given by the amplitude of the potential in the false vacuum, $V_0$. However, a better approximation for such a pre-exponential factor is given by the mass of the field in the meta-stable vacuum \cite{Linde:1981zj,Linde:2005ht}. We noticed that, upon such a replacement for the pre-exponential factor, the parameter space of models that exit inflation through a first order phase transition expands substantially. Having found the new parameter space, we calculate the amplitude of the gravitational radiation from such a first order phase transition. Like any gravitational wave profile which is generated from a first order phase transition, the spectrum has an inverted V-shape which is determined by the peak frequency and the amplitude at the peak frequency. The peak frequency  also depends on the reheating temperature after the first order phase transition, but if all the energy in the false vacuum phase is converted to radiation, the peak frequency falls somewhere in the GHz range, which is in the frequency range probed by Chongqing University \cite{Li:2003tv} but below its current sensitivity limit by $10^{-2}-10^{-5}$. Nonetheless, it is hoped that future improvements of the detectors improve their sensitivity and close the gap between the predicted signal and the current sensitivity limit. The shape of the produced gravitational spectrum can be used to distinguish the mechanism of termination of inflation from the parametric resonance  \cite{Kofman:1997yn} which leads to gravitational wave spectrum with different profile
\cite{Dufaux:2007pt, Ashoorioon:2013oha}. 

We also investigate the gravitational profile from inflationary models that exit from inflationary phase  through violation of slow-roll  but after the termination of inflation the nucleation rate from meta-stable vacuum to the true vacuum becomes large enough that first order phase transition occurs. If the reheating is instantaneous and efficient, the peak frequency from such phase transitions lies in the tera to peta-Herz band and its amplitude is much smaller than the models that exit from inflation through first order phase transition. At the moment there is no planned probe that aims to the detection of such high frequency gravitational waves.

In this work, we chose  the rolling potential to be the quadratic potential for simplicity. Also, the large-scale predictions of the model were very close to $m^2\phi^2$ prediction in the $n_S-r$ plane in the limit of vanishing vacuum energy. This part of the predictions of the model was still within the $2\sigma$ limit of Planck 2013 results \cite{Ade:2013uln}.  After the revelation  of Planck 2015 results \cite{Planck:2015xua}, the prediction of the model now falls outside the $2\sigma$ region in the $n_s-r$ plane excluding the running of scalar spectral index from the parameters. Nonetheless, the region close to the prediction point of  $m^2\phi^2$ is still within the $3\sigma$ region. Allowing for the running of scalar spectral index, this region comes back to the $2\sigma$ C.L. region.  It is expected that taking the rolling potential to be a low energy scale model, like the hilltop model \cite{Ashoorioon:2011ki} or the Starobinsky model \cite{Starobinsky:1980te}, the predicted values for $r$ for the models exiting inflation with a first order phase transition have a larger overlap with the $2\sigma$  region of Planck 2015 data.

The outline of the paper is as follows. First we will explain the setup that can realize the idea of double-field inflation. We identify the parameter space for which inflation can end through a first order phase transition and then calculate the power spectrum of gravitational waves produced during such phase transitions. Then we identify the region of parameter space where phase transition happens after the termination of inflation through slow-roll violation. We also compute the gravitational signature of such phase transitions. At the end, we show that the universe gets trapped in the meta-stable vacuum if the vacuum energy is less than a threshold. We comment on the viability of such inflationary models and then conclude our paper.

\section{Extended Hybrid Inflation}\label{The-setup}

We focus on the extension of hybrid Inflationary  potential \cite{Copeland:1994vg,Cortes:2009ej} given by \footnote{For the status of hybrid inflation with quadratic rolling potential after Planck 2013 data release, please see  \cite{Clesse:2014fwa}. In particular the paper discusses an interesting scenario in which the first part of the required number of e-folds, needed to solve the problems of Big Bang, is provided from the chaotic phase and the second part is resulted from the vacuum dominated phase of hybrid inflation. The model can render lower scalar spectral indices but larger amount of tensor-to-scalar ratio in comparison with the purely quadratic potential.}
\bea\label{Hybrid-Pot}
V(\phi,\psi)&=&\frac{1}{4}\lambda M^4+\frac{1}{2} m^2 \phi^2+  \nonumber\\
&&   \frac{1}{4}\lambda \psi^4-\frac{1}{3}\gamma M \psi^3 +\frac{1}{2} \lambda^{\prime} \phi^2\psi^2+\frac{1}{2} \alpha M^2 \psi^2
\label{Potential}
\eea
which is supplemented by the cubic term for the field $\psi$ to provide the possibility of first order phase transitions at the end of inflation. For large values of $\phi$, the potential has one minimum in both $\phi$ and $\psi$ direction. As inflation proceeds and $\phi$ rolls toward its vacuum, second minimum along the $\psi$ direction develops if $\gamma^2> 4\alpha \gamma$ at
\be
\phi_{\rm inflection}^2=M^2 \frac{\gamma^2-4\alpha \lambda}{4\lambda^{\prime}\lambda}
\ee
The two minima are separated by a barrier that allows for a first order phase transition from the meta-stable vacuum to the true one if the nucleation rate is substantial and when it is energetically favorable. Mapping the potential to \eqref{pot-tot}, $V_1(\phi)$ is the vanilla chaotic model $m^2\phi^2$ and the expression in the second line plays the role of $V_2(\phi,\psi)$, which couples the inflaton to the field that creates the false vacuum in the $\psi$ direction and facilitates the phase transition. In principle, $V_1(\phi)$ determines the predictions of the model  at large scales in the limiting case that the cosmological constant goes to zero and can be chosen such that the model is compatible with the CMB observables at cosmological scales \cite{Ade:2013uln}. 

The probability of phase transition is given by the nucleation rate divided by the 4-dimensional Hubble volume 
\begin{equation}
p=\frac{\Gamma}{H^4}\,.
\label{probability}
\end{equation}
where $\Gamma$ is the nucleation rate \cite{Callan:1977pt}
\be
\Gamma= {\cal A} \exp(-S_{\rm E})
\ee
where $S_{\rm E}$ is the  Euclidean four-dimensional action for the solution that interpolates between the two minima. For a first order phase transition with a quartic polynomial potential, the numerical results were fit by \cite{Adams:1993zs} to have the form
\begin{equation}
S_{\rm E}= \frac{4 \pi^2}{3 \lambda}(2-\delta)^{-3}(\alpha_1
\delta+\alpha_2 \delta^2+\alpha_3 \delta^3) \,,
\label{Euclidian-act}
\end{equation}
where $\alpha_1= 13.832 ,~\alpha_2=-10.819,~\alpha_3=2.0765$, and
$\delta$ is a function of $\phi^2$,
\begin{equation}\label{delta}
\delta=\frac{9 \lambda \alpha}{\gamma^2}+\frac{9 \lambda \lambda'
\phi^2}{\gamma^2 M^2} \,.
\end{equation}
The allowed range has $0<\delta<2$. Prefactor ${\cal A} $ has dimension mass$^4$. In \cite{Cortes:2009ej}, this prefactor is taken to be equal to the cosmological constant in the false vacuum, $\frac{\lambda M^4}{4}$. However as pointed out by Linde in \cite{Linde:1981zj}, if $M_{\psi}^2>2H^2$, the pre-exponential factor ${\cal A}$ is closer to $M_{\psi}^4$.  As we will see, upon this change, the parameter space in which exit from inflation can happen through a first order phase transition expands substantially\footnote{If $M_{\psi}^2<2 H^2$, the prefactor is of order $H^4$ \cite{Linde:2005ht}.}.

For illustration we focus on the following set of parameters 
\bea\label{parameters}
\alpha=0.01\,,\quad \lambda=1\,,\quad \& \quad \lambda^{\prime}=1\,.
\eea
All massive parameters are in unit of Planck mass, $m_{\rm P}=G^{-1/2}=1.2209\times 10^{19}~{\rm GeV}$. For the  potential \eqref{Potential} to have a zero vacuum constant after the phase transition, the following relation between the parameters should hold
\be
\gamma^2=\frac{-\alpha^3+9 \alpha \lambda^2 +(\alpha^2 +3\lambda^2)^{3/2}}{3}
\ee
As in \cite{Cortes:2009ej}, we take the required number of e-folds to solve the standard big bang cosmology problems to be $55$. Inflation can also end through end of slow-roll which is parameterized by
\begin{equation}
\epsilon \equiv \frac{m_{\rm Pl}^2}{16 \pi}
\left(\frac{V_1^{\prime}}{V_1}\right)^2 = \frac{m^4 \phi^2 \, m_{\rm
P}^2}{\pi (\lambda M^4 + 2 m^2 \phi^2)^2}= 1 \,.
\end{equation} 
which has the biggest solution
\begin{equation}\label{eps}
\phi^2_{\epsilon} = \frac{m^2 m_{\rm Pl}^2 + m m_{\rm P} \sqrt{m^2
    m_{\rm Pl}^2 -8\pi\lambda M^4}-4\pi\lambda M^4}
{8 \pi m^2} \,,
\end{equation}
as the physically viable solution.

In order for inflation to end through phase transition rather than slow-roll violation, $\phi_{\rm pt} > \phi_{\epsilon}$. The number of e-foldings could be obtained as a function of the inflaton field. As stated above, we take the required number of e-folds equal to 55:
\begin{equation}
N(\phi_{*},\phi_{\rm pt})= -\frac{8 \pi}{m_{\rm P}^2}
\int_{\phi_{*}}^{\phi_{\rm pt}}\frac{V_1}{V_1^{\prime}} \, d\phi =2 \pi \lambda \frac{M^4}{m^2 m_{\rm P}^2}
\ln \frac{\phi_{*}}{\phi_{\rm pt}} + \frac{2\pi}{m_{\rm
    P}^2}(\phi_{*} ^2 - \phi_{\rm pt}^2) \,
\end{equation}
From the above one can obtain $\phi_{*}$. The scalar spectral index and the tensor over scalar ratio then could be obtained at $\phi_{*}$ using 
\begin{eqnarray}\label{nS}
n_{\rm S}-1&=&-6 \epsilon + 2 \eta \,;\\
r&=& 16 \epsilon \,, \label{r}
\end{eqnarray}
where $\epsilon$ is given above and 
\begin{equation}\label{eta}
\eta \equiv \frac{m_{\rm P}^2}{8 \pi} \frac{V''}{V}
=\frac{m^2 \, m_{\rm P}^2}{2\pi (\lambda M^4 + 2 m^2 \phi^2)} \,.
\end{equation}

\section{End of Inflation with a first Order Phase Transition}\label{End-First-Order}
\begin{figure}[t]
  \includegraphics[scale=0.62]{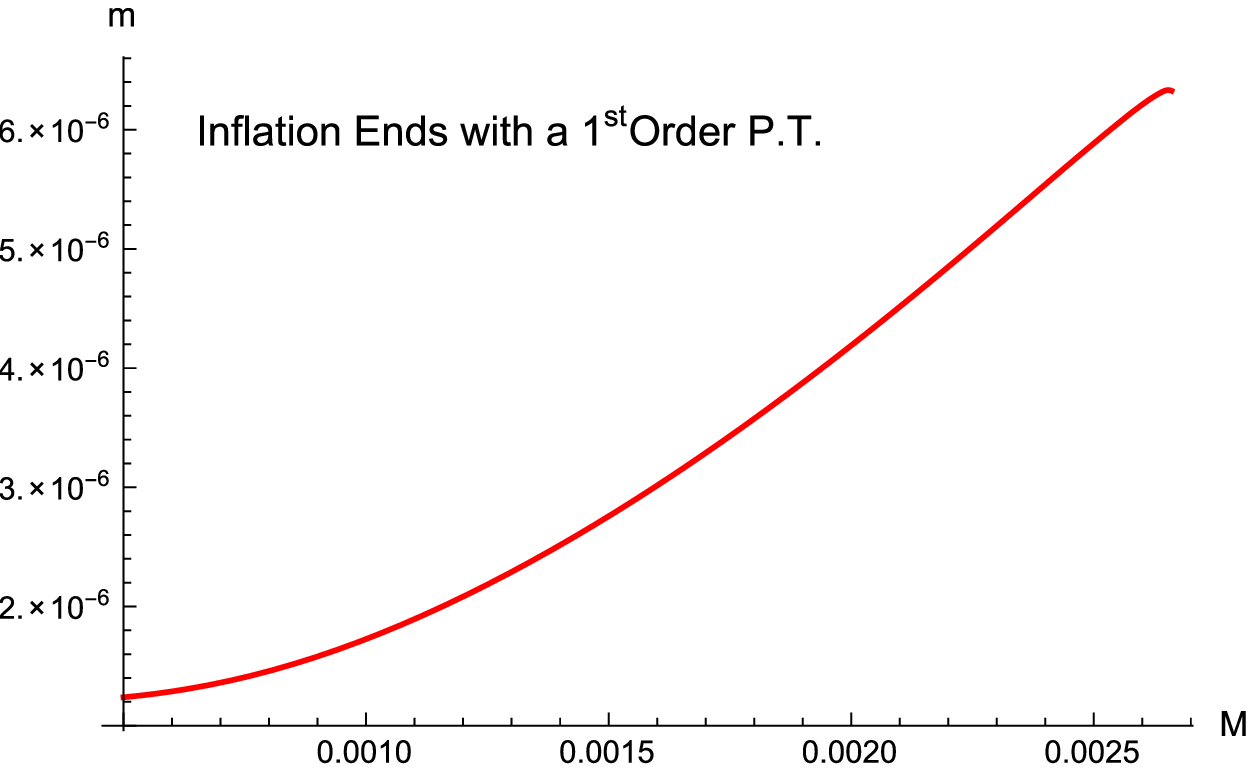}
  \includegraphics[scale=0.62]{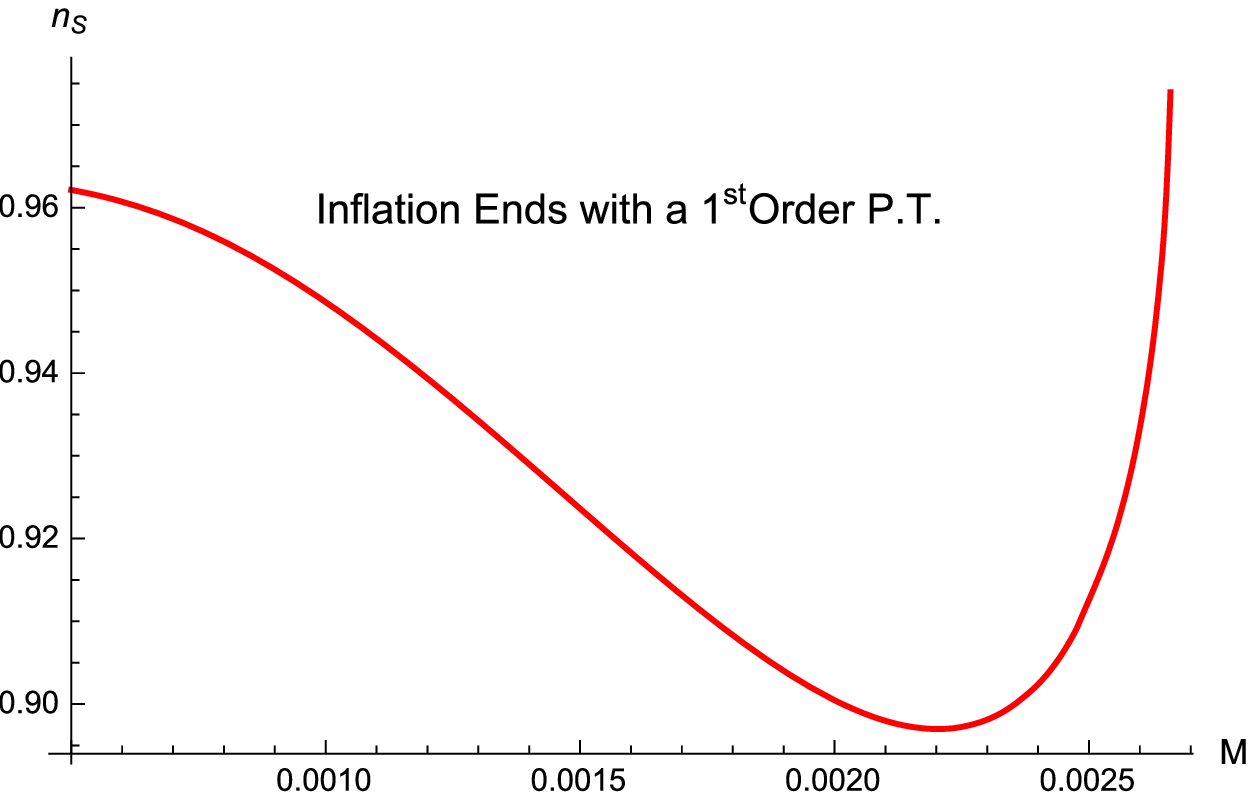}
  \caption{The behavior of $m$ (left) and $n_S$ (right) as a function of $M$ in the region of parameter space that inflation ends with a first order phase transition. }\label{m-and-ns-vs-M}
\end{figure}

\begin{figure}[t]
  \includegraphics[scale=0.62]{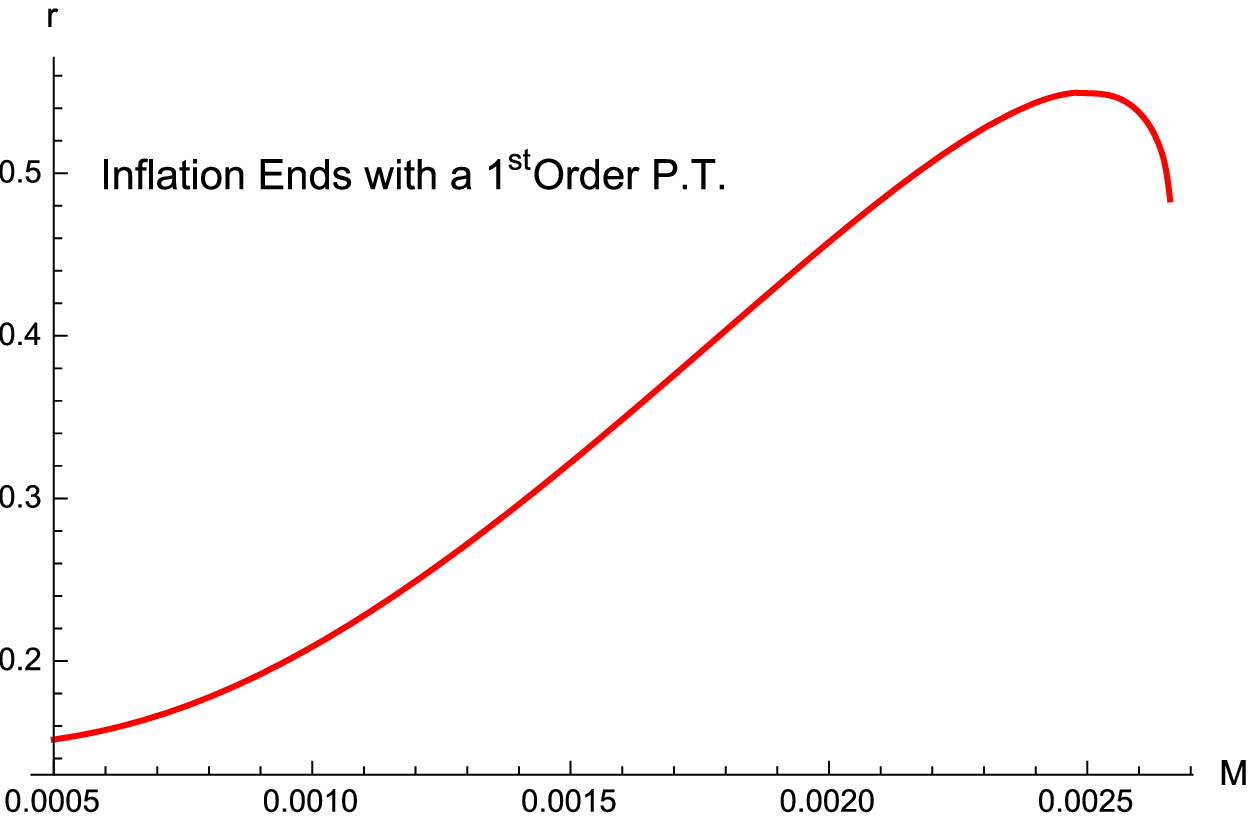}
  \includegraphics[scale=0.62]{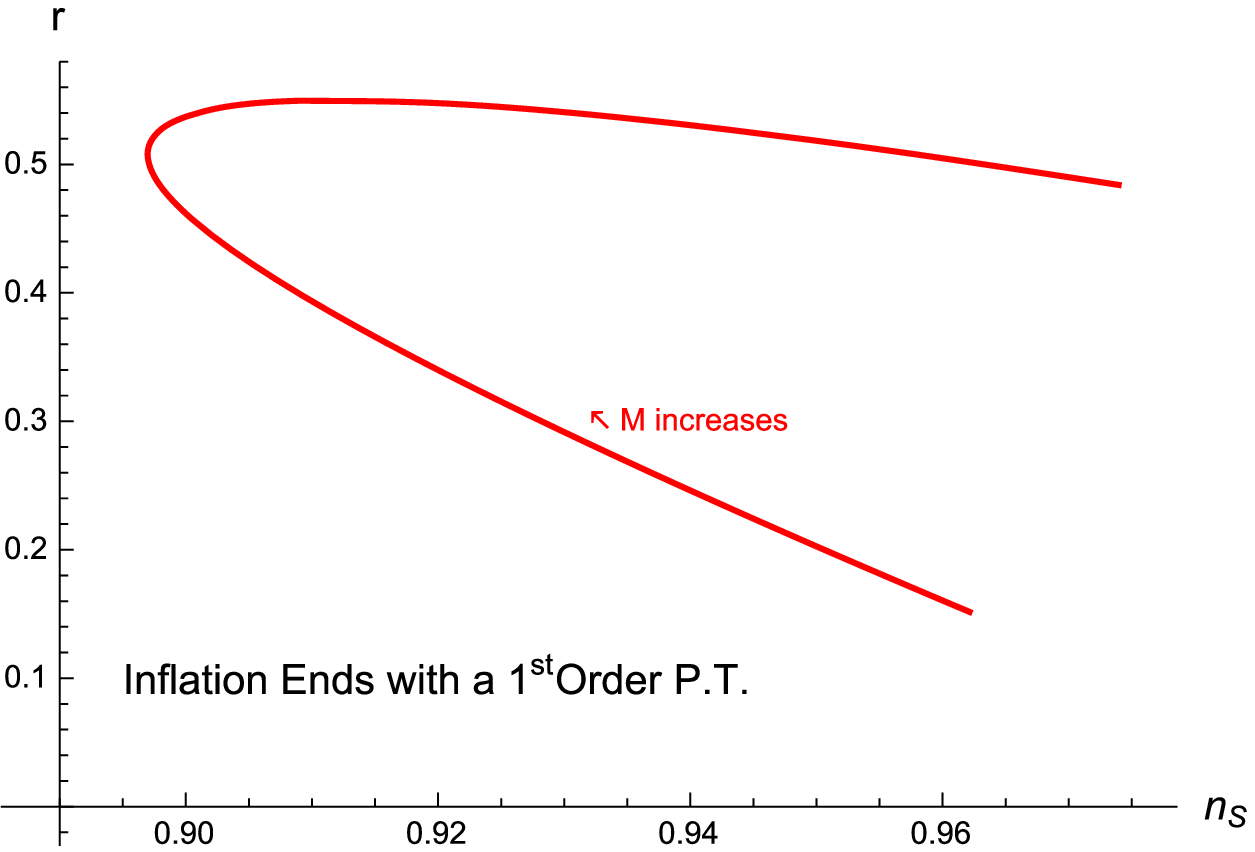}
  \caption{ The behavior of $r$ vs. M (left)  and vs. $n_S$ (right) in the region of parameter space that inflation ends with a first order phase transition.}\label{r-vs-M-and-ns}
\end{figure}

In this section we focus on the region of parameter space where before the violation of the slow-roll parameter, the nucleation rate becomes substantial enough that leads to percolation of true  vacuum bubbles in the sea of inflating false vacuum.  We assume that the pre-exponential factor, $\cal{A}$ is of order $M_{\psi}^4$, as in \cite{Linde:1981zj}, which makes our calculations distinct from the analysis of \cite{Cortes:2009ej}. As we will see, this assumption will expand the region of parameter space in which end of inflation happens through a first order phase transition. This part of parameter space is part of the region in which the vacuum energy contribution to the potential is comparable to the energy density of the inflaton field $\phi$. Detailed computation shows that only in the region where
\be\label{M-range}
4.97 \times 10^ {-4}\lesssim M \lesssim 2.66\times 10^{-3} \,, 
\ee
first order phase transition precedes the slow-roll violation.  For $M\gtrsim 2.66\times 10^{-6}$, there is no solution to the density perturbation amplitude normalization and hence no viable model.  One can match the amplitude of density perturbations with the COBE normalization which fixes the mass parameter of the scalar field, $m$.  In the left plot of fig.\ref{m-and-ns-vs-M}, we have graphed how the mass parameter varies as a function of $M$. The obtained range of $m$ is 
\be\label{m-range}
1.24 \times 10^ {-6}\lesssim m \lesssim 6.33\times 10^{-6} .
\ee
Within the range \eqref{M-range}, $n_S$ and $r$ vary in the following ranges
\bea\label{ns-r-range}
0.91\lesssim n_S \lesssim 0.97\,, \\
0.15\lesssim r \lesssim 0.55 \,,
\eea
as in the right graph in fig.\ref{m-and-ns-vs-M} and left plot of fig.\ref{r-vs-M-and-ns}. We have also plotted the predictions of this region of parameter space in the $n_S-r$ plane. As can be seen from the plots, with increasing $M$, initially $n_S$ decreases and $r$  increases. However there are turning points: around $M\simeq 2.2\times 10^{-3}$, $n_S$ starts to increase and around $M\simeq 2.4\times 10^{-3}$, $r$ starts to decrease. The loci of predictions of the model is designated in the $n_S-r$ plane in the right plot of fig.\ref{r-vs-M-and-ns}. For small values of $M$, the predictions of the models for large scale fluctuations is very close to the chaotic quadratic potential.

With the choice of parameters as in \eqref{parameters}, the true minimum in the $\psi$ direction appears very late, \ie towards the end of inflation. There is no minimum for $\psi$ that a bubble of true vacuum forms and lead to open inflationary scenario, as in \cite{Linde:1995rv}. 

If the energy stored in the inflaton potential is completely transformed to the radiation, \ie reheating is instantaneous, one can calculate the reheating temperature at the end of inflation  
\be\label{T-reheat}
T_{\ast}=\left(\frac{90 H_{f}^2}{8 \pi^3 g_{\ast}} \right).
\ee

where $g_{\ast}$ is the the total number of relativistic degrees of freedom at temperature  $T_{\ast}$ which we take to be $g_{\ast}\simeq 106$. We have plotted $T_{\ast}$ vs. $M$ in the left plot of fig.\ref{T-and-beta-vs-M}. If reheating is not efficient,  $T_{\ast}$ is smaller than what is given in \eqref{T-reheat}. This temperature determines the peak frequency of the gravitational wave spectrum generated through the first order phase transition.  If one assumes instant reheating, the reheating temperature varies almost linearly in the range $10^{15}$ GeV and $10^{16}$ GeV  vs.  $M$ in the range  \eqref{M-range}.

\subsection{Gravitational Wave Spectrum from Models with First Order Phase Transition at the end of Inflation}

\begin{figure}[t]
  \includegraphics[scale=0.62]{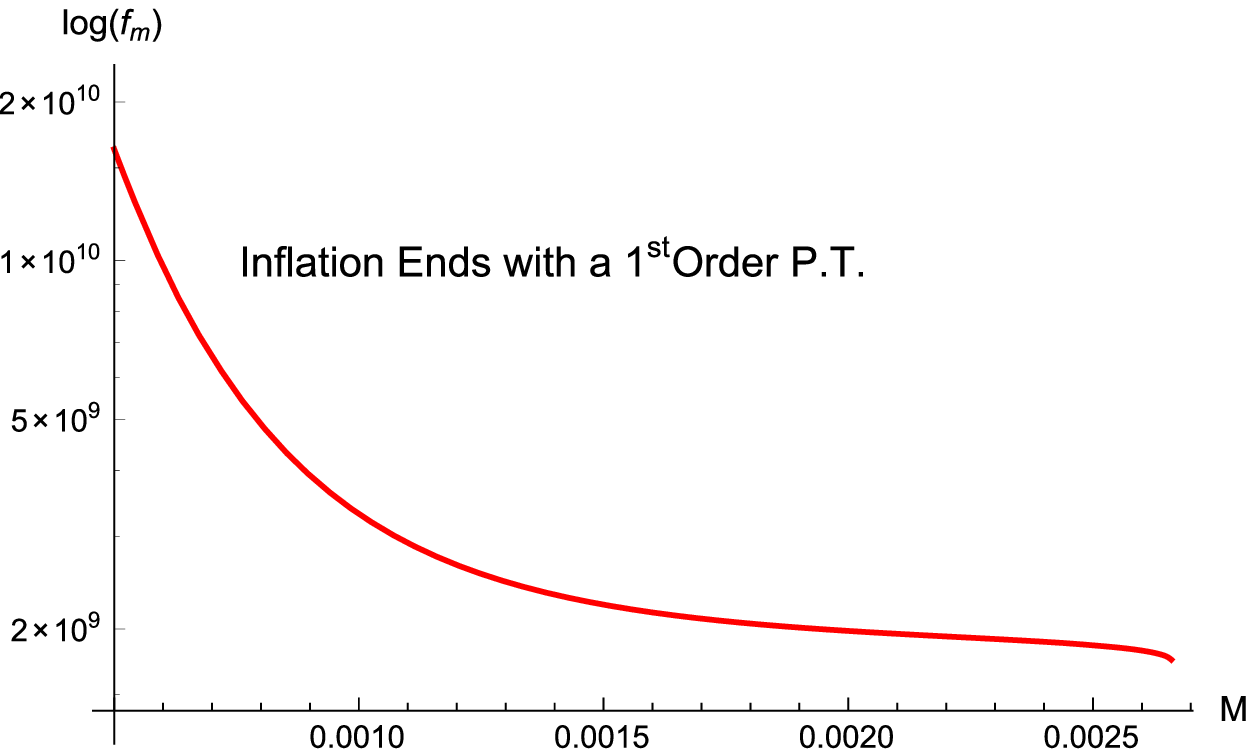}
  \includegraphics[scale=0.62]{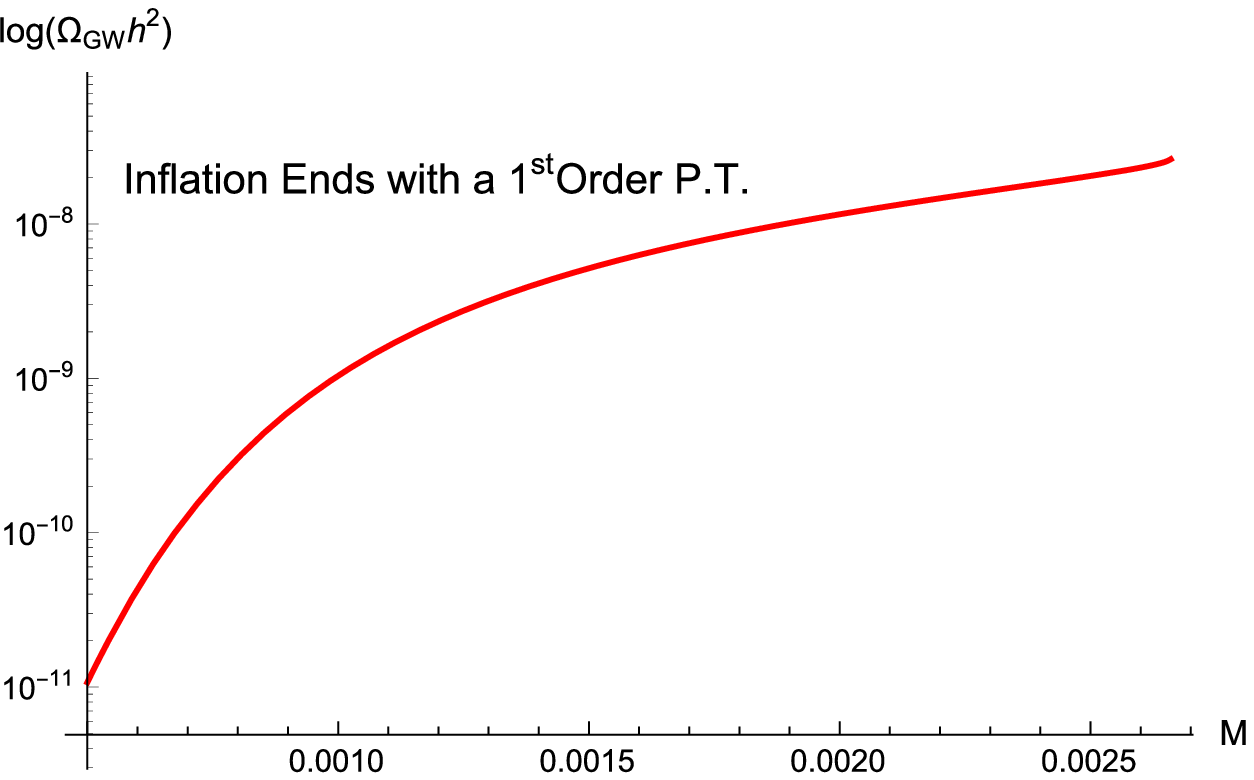}
  \caption{The peak frequency of the gravitational wave spectrum and its amplitude as a function of $M$ for models that exit inflation with a first order phase transition. }\label{fm-and-I-vs-M}
\end{figure}

Since inflation supercools the universe, one can exploit the formalism of gravitational wave generation from first order phase transition at zero temperature. The numerical computations for bubble collisions from a first order phase transition was initially done by \cite{Kosowsky:1992vn} for two bubbles and generalized for more bubbles in \cite{Huber:2007vva}.  The spectrum has the shape of asymmetric of inverted V around a peak frequency, $f_{\rm m}$ , which decays  like $f^{2.8}$ and $f^{-1}$ respectively at smaller and larger frequencies. The peak frequency $f_m$ today, after the post-inflationary redshifting is taken into account, is given by \cite{Kosowsky:1992vn}
\begin{figure}[t]
  \includegraphics[scale=0.62]{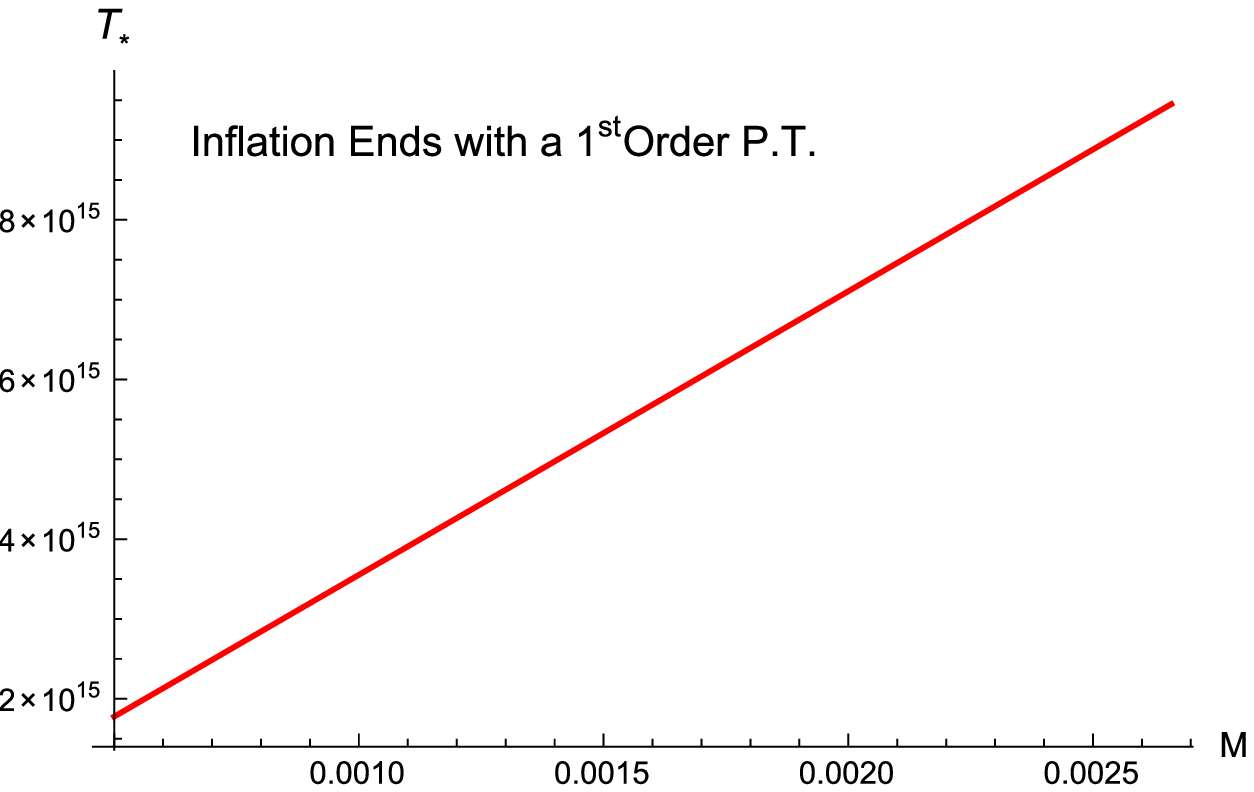}
  \includegraphics[scale=0.62]{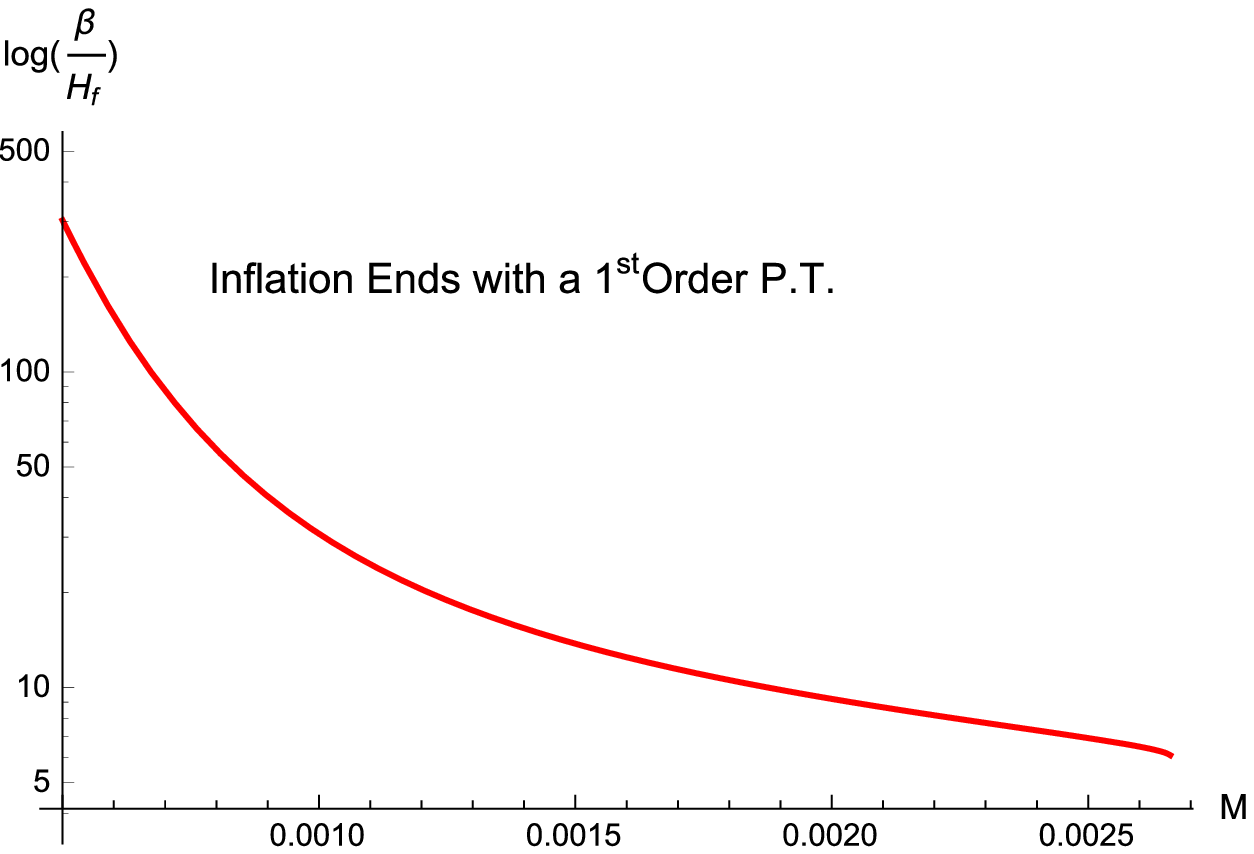}
  \caption{ The behavior of $T_{\ast}$ and $\left(\frac{\beta}{H_f}\right)$ vs. $M$ in the region of parameter space that inflation end with a first order phase transition.}\label{T-and-beta-vs-M}
\end{figure}

\be\label{frequency}
f_{\rm m}=3\times 10^{-10} {\left(\frac{g_{\ast}}{100}\right)}^{1/6}  \left(\frac{T_{\ast}}{1  \rm GeV}\right) \left(\frac{\beta}{H_f}\right),
\ee 
and the amplitude at this peak frequency is given by
\be\label{GW-intensity}
\Omega_{\rm GW} h^2 (f_m)=10^{-6} \left(\frac{g_{\ast}}{100}\right)^{1/3} {\left(\frac{H_f}{\beta}\right)}^2,
\ee
where $\beta$ is 
\be
\beta=\frac{dS_{\rm E}}{dt}= \frac{dS_{\rm E}} {d\phi} \frac{d\phi} {dt}.
\ee
$\beta^{-1}$ is a measure of how fast the first order phase transition takes to complete. For our computations to be reliable, we expect this time to be much smaller than the expansion rate of the universe, \ie  $\beta/H_f\gg 1$, where $H_f$ is the Hubble parameter at the time of phase transition which coincides with the end of inflation. $ \frac{d\phi} {dt}$ is the velocity of the scalar field, $\phi$, which can be found during inflation through the following relation \cite{Ashoorioon:2004wd}
\be
{\dot{\phi}}^2=\frac{2\epsilon(\phi) V_1(\phi)}{3-\epsilon(\phi)}
\ee
where $\epsilon$ is the first slow-roll parameter. We have plotted $\log\left(\frac{\beta}{H_f}\right)$ vs. $M$ in the range  \eqref{M-range}  of $M$ that leads to exit from inflation with a first order phase transition. As can be seen from the right plot of fig.\ref{T-and-beta-vs-M}, for smaller values of $M$ in the range, phase transition is quite fast in comparison with the expansion rate of the universe, $\beta/H_f\sim {\rm few}\times 100 $.  As $M$ increases, phase transition takes longer to complete in comparison with the expansion time of the universe. For the maximum value of $M$ in the range, $M=2.66\times 10^{-3}$,  $\beta/H_f=6.11$, which is nonetheless fast enough to validate our computations. The intensity of the gravitational waves at the peak frequency is only dependent on this parameter, $\beta/H_f$. The faster the phase transition, the smaller the amplitude of the produced gravitational waves.  We have plotted $\log(\Omega_{\rm GW} h^2)$ vs. $M$ which increases in the range  \eqref{M-range} from $1.07\times 10^{-11}$ to $2.63\times 10^{-8}$, see the right plot in fig.\ref{fm-and-I-vs-M}. However the peak frequency, besides  $\beta/H_f$ depends on the reheating temperature too. Assuming that the reheating is instantaneous and the total energy density of potential at the end of inflation transforms to radiation, we have  plotted the  peak frequency as a function of $M$ in the range \eqref{M-range}, see left plot of fig.\ref{fm-and-I-vs-M}. For the smallest value of $M$ in the range, $M\simeq 5\times 10^{-4}$, peak frequency, $f_{\rm m}$ is around $1.63 \times 10^{10}$ Hz whereas for the largest value of $M$, $f_{\rm m}\simeq 1.75\times 10^9$ Hz. Such a gravitational wave profiles lies in the frequency band of Chongqing High Freqeuncy Gravitational (HFGW) probe but below its current sensitivity limit by a factor of $10^{-2}-10^{-5}$ \cite{Li:2003tv}. It is hoped that the improvement of the detector in future close the gap between the expected signals and the current sensitivity limit.

 If the reheating coming from bubble wall collision is not instantaneous and a phase of non-radiation domination intervenes the end of first order phase transition and the radiation domination, $T_{\ast}$  will be lowered. If this efficiency factor is assumed to be $\chi$, where $\chi\lesssim1$, the amplitude at the peak frequency will be lowered by a factor of $\chi^4$ \cite{Lopez:2013mqa}. For example if  $\chi\simeq 0.01-0.1$, the frequency range will be shifted within the sensitivity band of Birmingham HFGW probe \cite{Cruise:2006zt}. If the reheating temperature is a factor of $\chi\simeq 10^{-7}-10^{-10}$ smaller, the peak frequency will shift to the sensitivity band of DECIGO \cite{Kawamura:2011zz} and BBO \cite{Crowder:2005nr}. However in all these cases, the amplitude of the signal will be lowered such that the signal could not be observed by any of these probes. 

With the choice of $V_1(\phi)$ as the quadratic potential, most of the parameter space of inflationary models that exit from inflation with a first order phase transition is ruled out. There is a small region of parameter space which has predictions very close to $m^2\phi^2$. This region in particular is of interest for the CMB polarization probes like the  BICEP2  \cite{Ade:2014xna} or the future ones like the CMBPol \cite{Baumann:2008aq}. Since these models exit inflation through first order phase transition which is accompanied by bubble collision, they can leave an extra signature of gravity waves at higher frequency scales.  We expect that the predictions of the model for $r$ could be lowered if one would replace $V_1$ with another model, like the hilltop model, \cite{Ashoorioon:2011ki}, or models with smaller energy scales that naturally predict a lower value for $r$. 

\begin{figure}[t]
  \includegraphics[scale=0.62]{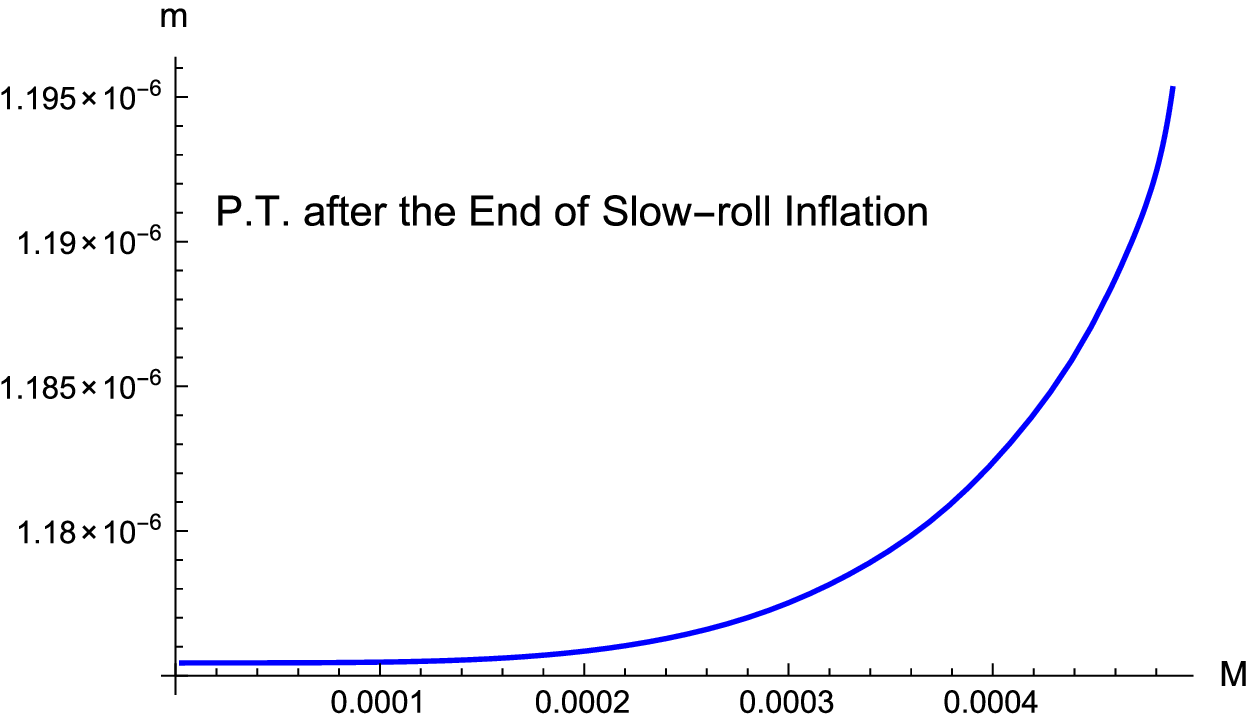}
  \includegraphics[scale=0.62]{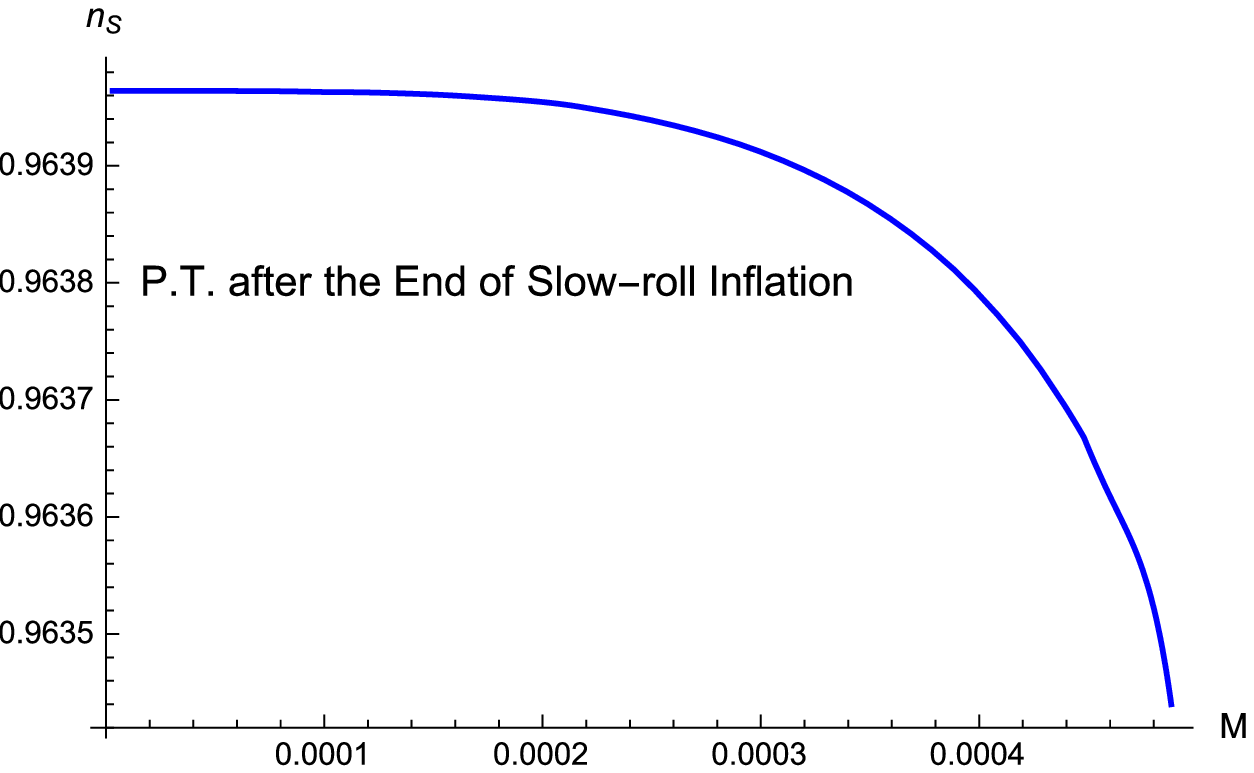}
  \caption{$m$ (left) and $n_S$  (right) vs. $M$ for the models for which first order phase transition happens after the end of inflation. }\label{m-and-ns-vs-M-after-pt}
\end{figure}

\section{First Order Phase Transition after the End of Slow-roll Inflation}\label{End-slow-roll}

\begin{figure}[t]
  \includegraphics[scale=0.62]{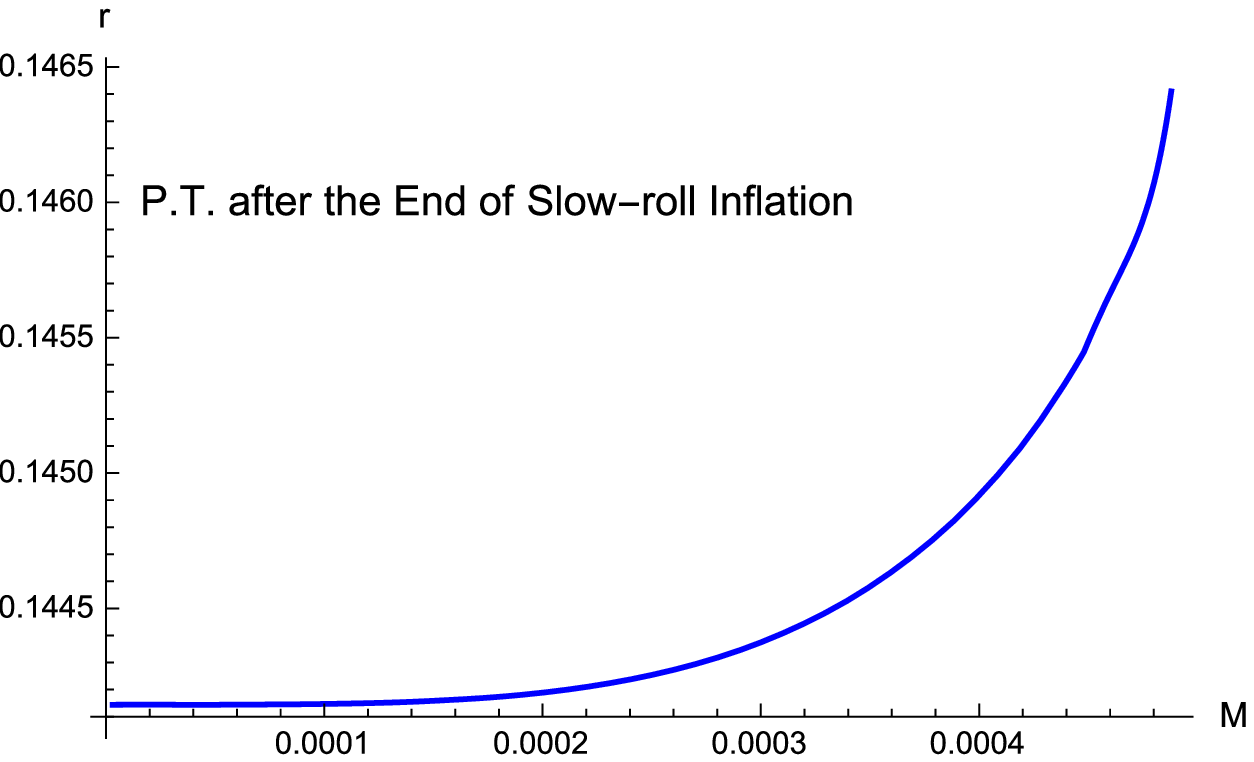}
  \includegraphics[scale=0.62]{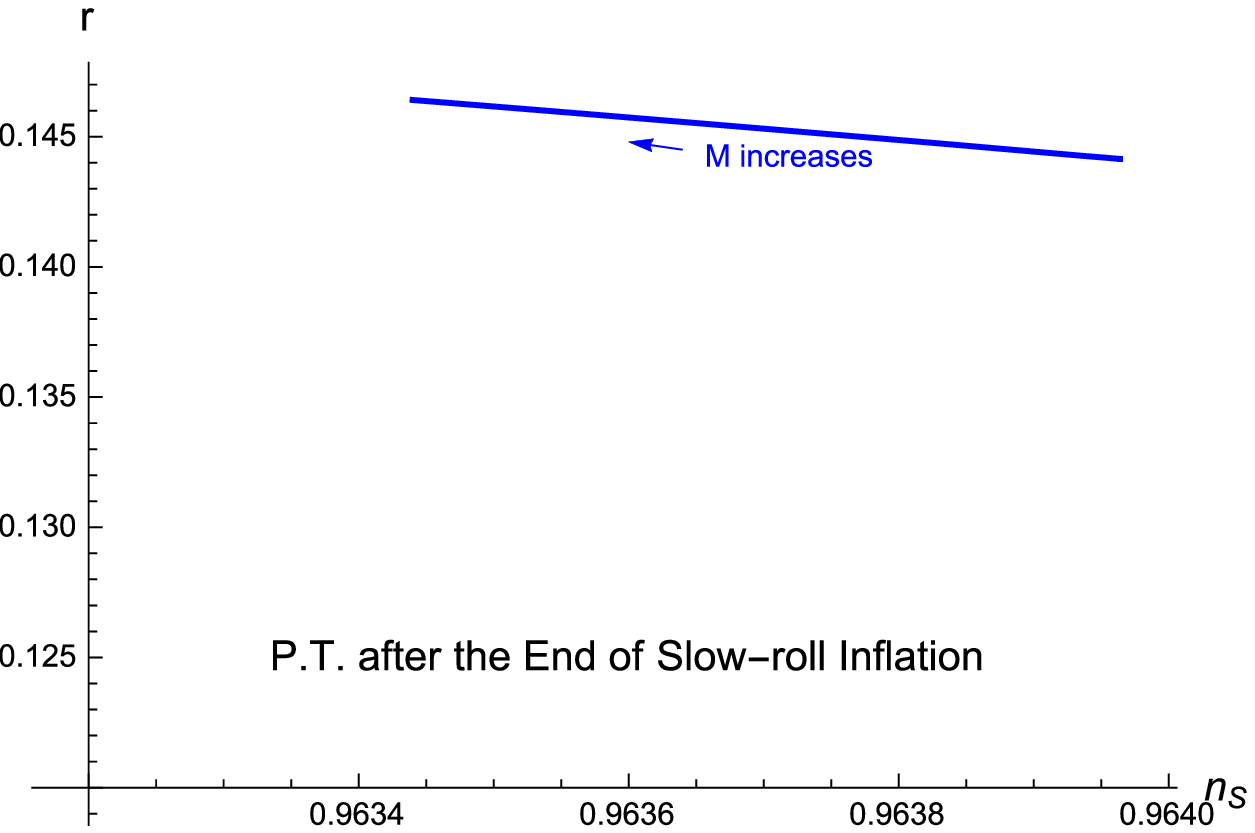}
  \caption{ $r$ vs. $M$ (left) and $n_S$ (right) for the models for which first order phase transition happens after the end of inflation. }\label{r-vs-M-and-ns-after-pt}
  \end{figure}

\begin{figure}[b]
  \includegraphics[scale=0.62]{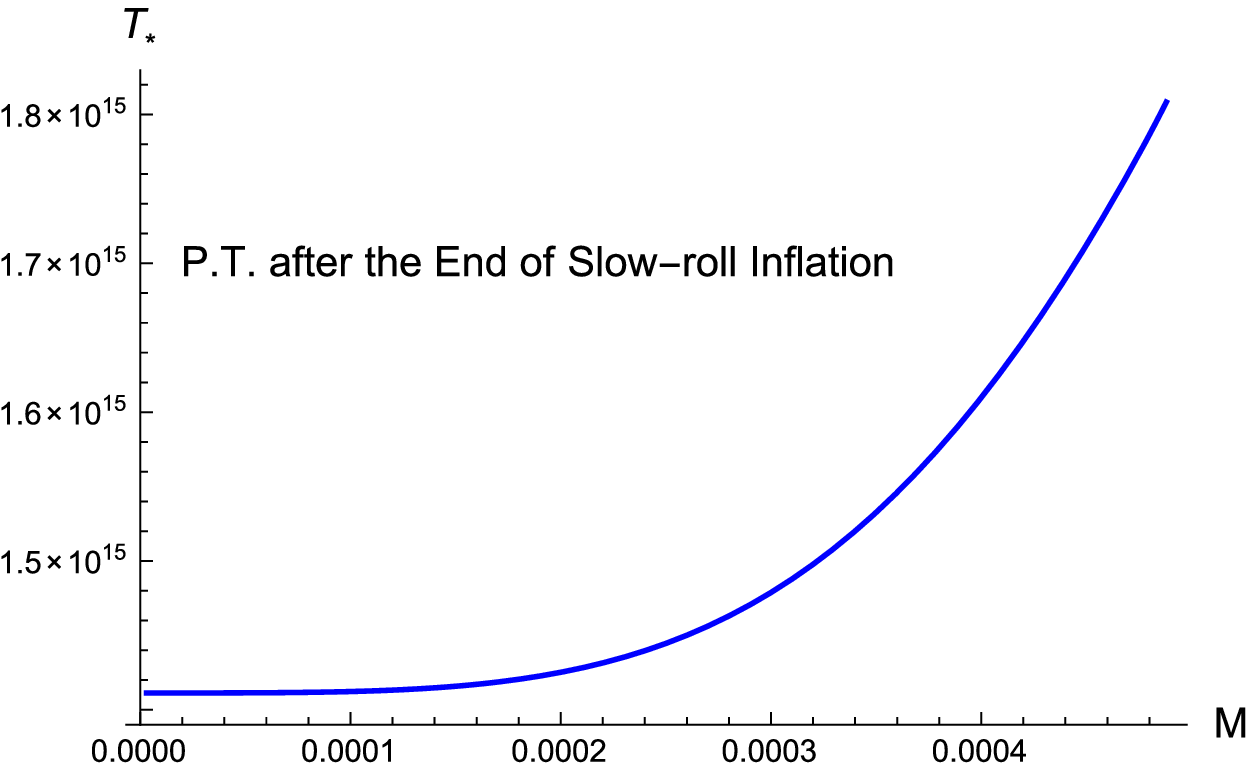}
  \includegraphics[scale=0.62]{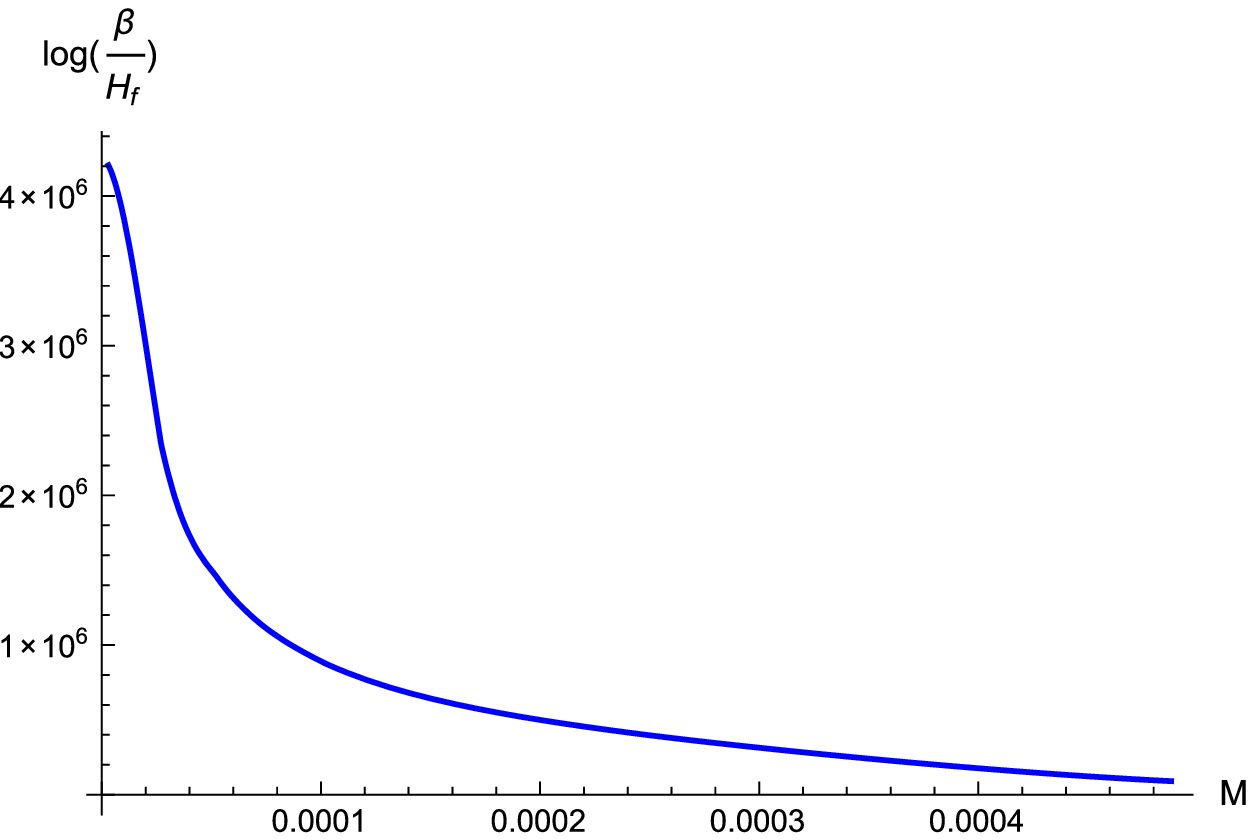}
  \caption{ $T_{\ast}$ (left) and $\log\left({\beta}{H_f}\right)$ (right) vs. $M$  for the models for which first order phase transition happens after the end of inflation. }\label{T-and-beta-vs-M-after-pt}
\end{figure}

It is possible that even though inflation ends through violation of the slow-roll inflation, the first order phase transition happens after the end of inflation, when $\phi$ becomes closer to the meta-stable vacuum, $\phi=0$. In fact if 
\be
3\times 10^{-6}\lesssim M\lesssim 4.88 \times 10^{-4}
\ee
phase transition happens after the end of slow-roll inflation.  The lower bound of the above interval is intriguingly very close to the scalar field mass in the rolling direction. To be able to determine this range of $M$, we solved the equations of motion and computed the evolution of the scalar field after inflation numerically\footnote{ In the range 
\be\label{M-range-PT-after-inf}
4.88 \times 10^{-4} \lesssim  M  \lesssim 4.97 \times 10^ {-4},
\ee
the constraints on the number of e-folds, $N_e=55$, with end of inflation given by the slow-roll violation did not yield a real solution for $m$. One can satisfy the constraint equations with less number of e-folds though.}. 
As before, we have to pick up only the $\phi$ solution to the nucleation rate equation that $0<\delta<2$. We have plotted the variation of $m$ as a function of $M$ in the range \eqref{M-range-PT-after-inf}, see the left plot in fig.\ref{m-and-ns-vs-M-after-pt}. Assuming that the reheating is instantaneous, we have also plotted the reheating temperature in this range of $M$, see the right plot in fig.\ref{m-and-ns-vs-M-after-pt}. In both cases, parameters $m$ and $T_{\ast}$ are almost constant for smaller values of $M$  in the range, but they gradually rise as $M$ increases. As expected, contrary to the behavior of $r$, $n_S$ decreases with the rise of $M$, please see fig. \ref{m-and-ns-vs-M-after-pt} and \ref{r-vs-M-and-ns-after-pt}. We have also plotted the behavior of $n_S$ vs. $r$ as $M$  increases, see fig.\ref{r-vs-M-and-ns-after-pt}. 
The behavior of $\log\left(\frac{\beta}{H_f}\right)$ vs. $M$, where $H_f$ is the Hubble parameter when phase transition completes can be seen in right plot in fig. \ref{T-and-beta-vs-M-after-pt}. As can be seen from the plot, with the increase of $M$, phase transition becomes slower. Nonetheless, in general, phase transitions happen much faster after inflation in comparison with the inflationary models that end via first order phase transition. This is partly due to the fact that the Hubble parameter after inflation ends,  is smaller than its corresponding value at the end of inflation. As $M$ enhances in the interval, \eqref{M-range-PT-after-inf}, $\beta/H_f$ decreases from $4.21\times 10^6$ to $90321.6$.  We have checked that in this range of $M$, the effective mass of $\psi$ field, $M_{\psi}^2>2H^2$ and thus phase transition happens via Coleman-de Luccia instanton transitions 
\cite{Callan:1977pt}. 

\begin{figure}[t]
  \includegraphics[scale=0.62]{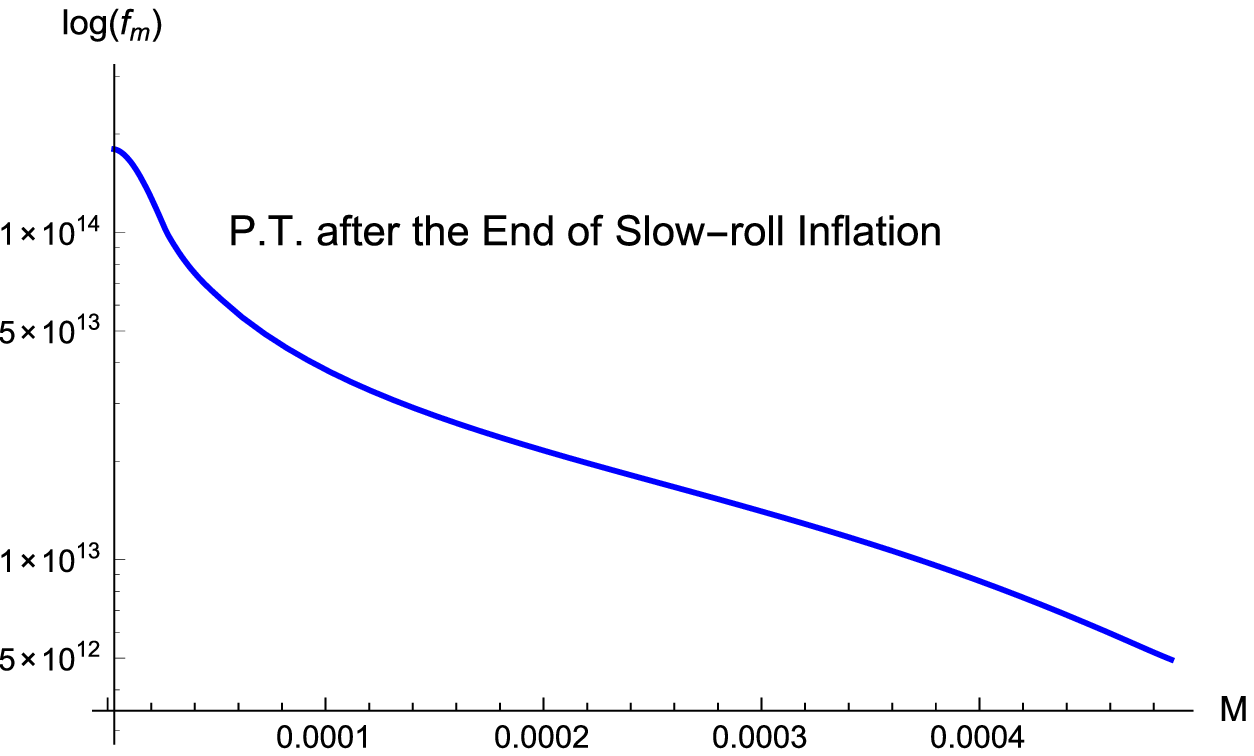}
  \includegraphics[scale=0.62]{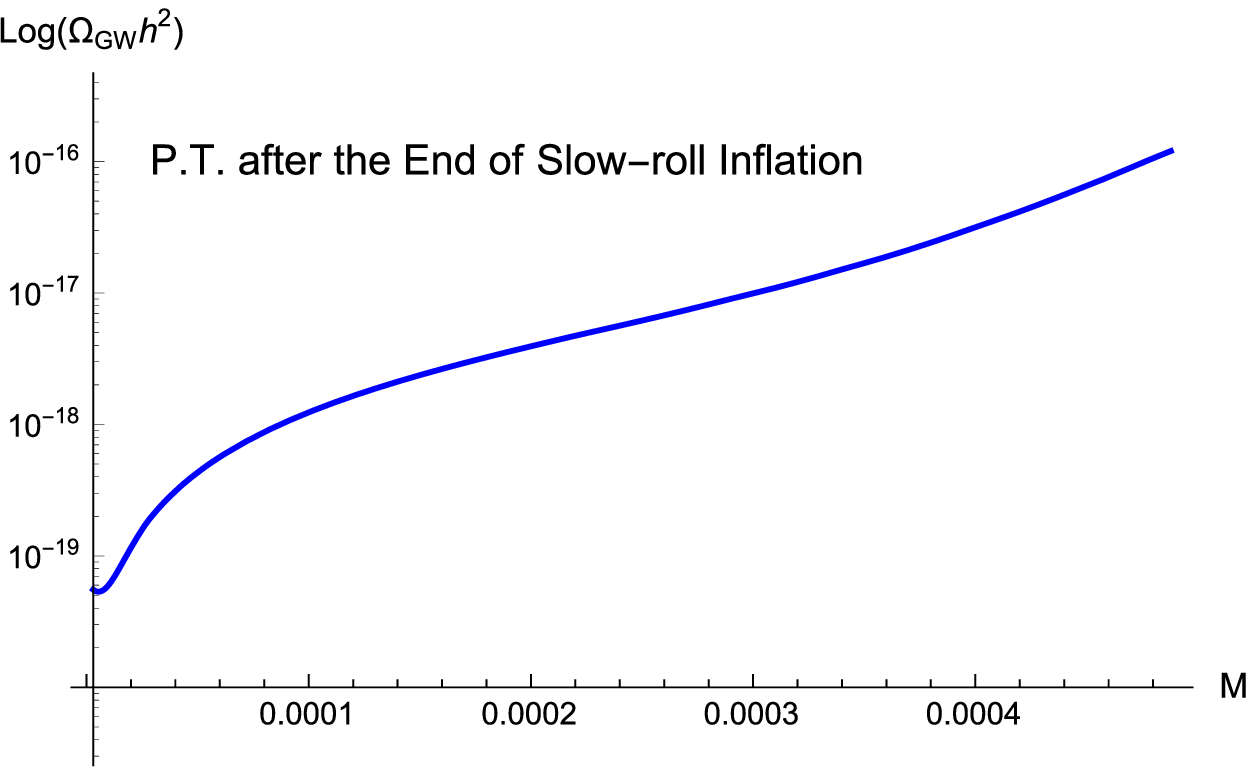}
  \caption{ The peak frequency (left) and the amplitude at the peak frequency (right) vs. $M$ for the models of inflation in which first order phase transition happens after inflation.}\label{fm-and-I-vs-M-after-pt}
\end{figure}

Using Eqs. \eqref{frequency} and \eqref{GW-intensity}, we have calculated the peak frequency of the gravitational wave spectrum and its intensity, $\Omega_{\rm GW} h^2$, as a function of $M$, please see the plots in fig.\ref{fm-and-I-vs-M-after-pt}.  As $M$ increases, $f_m$ varies from $1.8\times 10^{14}$ to $4.95\times 10^{12}$ Hz, which is well outside the frequency band of any currently planned probe. Of course this frequency range is obtained assuming that the reheating is instantaneous and efficient. If the reheating temperature is smaller than its instantaneous value by a factor of $\chi\simeq 10^{-3}-10^{-4}$, the peak frequency range is moved to the sensitivity band of Chongqing high frequency gravitational wave probe. However, the intensity will then gets suppressed further by a factor of $10^{-12}-10^{-16}$ which makes the signal too small to be detected.  

\section{Trapping in the Meta-Stable Vacuum and Eternal Inflation}\label{Meta-Stable vacuum}

For smaller values of $M$, $M< 3\times 10^{-6}$, the nucleation rate as given by Coleman-De Luccia transition rate, is too small to allow for the first order phase transition to complete as the scalar field, $\phi$, passes through the minimum of the potential, $V_1$. Gradually as the effective mass of the $\psi$ field decreases while the $\phi$ field approaches the metastable vacuum, there is a chance that $M_{\psi}^2\leq 2H^2$. In particular this can happen for smaller values of $\alpha$ and $M$. In this situation, Hawking-Moss phase transition \cite{Hawking:1981fz} appears where an inhomogeneous bubbles whose radius is greater than de-Sitter space radius, $H^{-1}$, will form in the flat space-time. The tunneling probability per unite four volume is of order  \cite{Hawking:1981fz}
\be
\Gamma= H^4 \exp(-B),
\ee
where 
\be
B=\frac{1}{8}\left[\frac{1}{V(0,0)}-\frac{1}{V(0,\psi_{\rm max})}\right]
\ee
is the difference between the combined gravitational and scalar field action of the $\psi=\psi_{\rm max}$, which is another homogeneous solution apart from $\psi=0$. $V(0,\Psi_{\rm max})$ is the local maximum of $V$ on a four-sphere of radius, $H_{\rm max}^{-1}$, where $H_{\rm max}^2=8\pi V(0,\psi_{\rm max})/3$. As $\psi=\psi_{\rm max}$ is unstable, the scalar field runs downhill from $\psi_{\rm max}$,  to the global minimum $\psi_{\rm min}$, afterwards. For the potential \eqref{Potential}, one can calculate the exponent $B$ and show that 
\be
B\simeq \frac{\alpha^3}{3\lambda^2 M^4}
\ee
for $\alpha\ll 1$. Therefore for
\be
M\lesssim \frac{\alpha^{3/4}}{3^{1/4}\lambda^{1/2}}
\ee
the Hawking-Moss phase transition rate from the meta-stable vacuum to the the true vacuum is very small. In our example where we took, $\lambda=1$ and $\alpha=0.01$, for $M\lesssim 6.08\times 10^{-5}$ such phase transition takes a lot of time to complete and basically leads to a self-reproduction regime like old inflation \cite{Guth:1980zm}. After the Hawking-Moss transition completes, the $\phi$ field will be homogeneous on scalar of order $H_{0}^{-1}$, with $H_0=\sqrt{\frac{8 \pi V(0,0)}{3 m_{\rm P}^2 }}$. Nonetheless, it will be stochastically distributed with the dispersion \cite{Linde:2005ht}
\be
\phi_{\rm rms}^2\equiv \langle \phi^2\rangle=\frac{3H_0^4}{8\pi^2m^4} ,
\ee
which in our example is about $\simeq 10^{-5}$ and thus too small to drive another phase of inflation.

In \cite{Cortes:2009ej}, the authors claim that they have been able to find a branch that corresponds to the vacuum-dominated regime of hybrid Inflation \cite{Copeland:1994vg} in the Extended Hybrid inflation, potential \eqref{Hybrid-Pot}. However in hybrid inflation, this regime is obtained assuming that the tachyonic instability in the waterfall field ends inflation. In this case, however the waterfield  mass squared, $\alpha M^2+\lambda \psi^2$ is always positive and never becomes tachyonic. For such energy scales in the vacuum dominated regime, the slow-roll can never get violated as the first slow-roll parameter decreases as inflation progresses. We also showed  that Hawking-Moss phase transition for such small vacuum energies is not substantial enough to end inflation. Therefore, it is not possible to realize the vacuum energy dominated regime of hybrid inflation in such extended models. The scalar field gets trapped in the metastable vacuum with no graceful exit and therefore this region of parameter space does not yield a viable inflationary model. On the other hand, from the phenomenological perspective, if $M$ is of order $\simeq 10^{-30} $, this vacuum energy can be responsible for the current acceleration of the universe. 

\section{Conclusion}

Terminating inflation with a first order phase transition has the benefit of reheating the universe from the supercooling phase of inflation through the collisions of bubbles of true vacuum without invoking and fine-tuning of the couplings of the inflaton to the other fields. One can achieve this scenario, modifying the old inflation scenario, with time-dependent nucleation rate which is small in the beginning and becomes substantial at the end of inflation. One specific realization of this scenario is extended inflation \cite{La:1989za}, in which the gravity sector of the theory is modified to  Jordan-Brans-Dicke theory. Another way of achieving this scenario, is having two scalar fields, where one of the fields rolls and the other one is trapped in a meta-stable vacuum \cite{Linde:1990gz,Adams:1990ds}. As the rolling field evolves, the nucleation rate at the false vacuum becomes large enough that the condition for percolation of true vacuum bubbles holds and inflation ends. We re-examined a model of extended hybrid inflation which provides us with such a setup. We noticed that the pre-exponential factor in the nucleation rate plays a crucial role in correctly determining the parameter space that allows for a first order phase transition in the model. For models that exit inflation with a first order phase transition, we computed the peak frequency and the amplitude at the peak frequency, which are respectively in the ranges $f_m\simeq 10^9-10^{10}$ Hz and $\Omega_{\rm GW} h^2 \simeq 10^{-11}-{10}^{-8}$. The signature is in the frequency range of Chongqing HFGW probe but smaller than the current sensitivity limit by a factor of $10^{-2}-10^{-5}$. It is expected that future improvements of the detectors of the probe can close this gap between the signal predicted by the model and the sensitivity limit \cite{Li:2003tv}. One should note that the predicted amplitude for the high frequency gravitational signal is the same as the gravitational signal generated by parametric resonance in the GUT scale models of inflation \cite{Easther:2006gt,Easther:2006vd}, although the two spectra could be distinguished from each other by their shape. In particular the gravitational wave profile generated by the parametric resonance has a double-hump feature instead of the single peak profile  of the gravitational wave signal generated by the bubble collision. 

We also showed that it is possible for some models to make a first order phase transition to the true vacuum after the termination of inflation with violation of the slow-roll parameter. However the signal of such a phase transition is weak and out of the frequency band of future probes.

The large-scale predictions of the model are very dependent on the potential of the rolling field. Here with the choice of quadratic potential for the rolling field, the predictions of the model approaches the predictions of $m^2\phi^2$ in the $n_S-r$ plane, which was still within the $2\sigma$ limit of Planck 2013 results \cite{Ade:2013uln}. As this model produces gravitational waves at the CMB scales, the analyzed model produces an observable gravitational waves at both small and high frequency range of the spectrum: the small frequency gravitational waves have quantum origin but the high frequency gravitational waves have classical origin. After the exposure of Planck 2015 results \cite{Planck:2015xua}, major prediction of the model is now outside the $2\sigma$ region in the $n_s-r$ plane excluding the running of scalar spectral index from the parameters. The region close to the prediction point of  $m^2\phi^2$ is still within the $3\sigma$ region though. If one allows for the running of scalar spectral index, this region comes back to the $2\sigma$ confidence region.  One should also note that the predictions of the model at large scales is very sensitive to the initial condition for fluctuations. As it was shown in \cite{Ashoorioon:2013eia}, choosing excited initial condition for cosmological perturbations generally tend to suppress the tensor-to-scalar ratio. 

It is expected that if one takes the rolling field potential to be a lower energy scale model, like the hilltop \cite{Ashoorioon:2011ki} or the Starobinsky \cite{Starobinsky:1980te} model, the predicted values for $r$ for the models exiting inflation with a first order phase transition have a larger overlap with the $2\sigma$ confidence region of Planck 2015 results. This is something that I will postpone to a future publication.

As the peak frequency of the gravitational spectrum, produced from the first order phase transition, depends very much on the Hubble parameter at the time of phase transition, it is also interesting to investigate inflationary models that exit inflation with smaller Hubble parameter and, hence, peak frequency that falls within the sensitivity bands of BBO, DECIGO or even Advanced LIGO.

\section*{Acknowledgements}

I am indebted to Andrei Linde for helpful discussions. I also thank  A. Abolhasani,  R. Allahverdi, M. Cortes, K. Freese and  A. Lopez for comments and discussions.


\begin{thebibliography}{99}

\bibitem{Guth:1980zm} 
  A.~H.~Guth,
  Phys.\ Rev.\ D {\bf 23}, 347 (1981).
  
\bibitem{Linde:1981mu} 
  A.~D.~Linde,
  Phys.\ Lett.\ B {\bf 108}, 389 (1982);
  A.~Albrecht and P.~J.~Steinhardt,
  Phys.\ Rev.\ Lett.\  {\bf 48}, 1220 (1982).
  
\bibitem{Kofman:1997yn} 
  L.~Kofman, A.~D.~Linde and A.~A.~Starobinsky,
  Phys.\ Rev.\ D {\bf 56}, 3258 (1997)
  [hep-ph/9704452];
  M.~A.~Amin, M.~P.~Hertzberg, D.~I.~Kaiser and J.~Karouby,
  Int.\ J.\ Mod.\ Phys.\ D {\bf 24}, 1530003 (2015)
  [arXiv:1410.3808 [hep-ph]].
  
\bibitem{Linde:1990gz} 
  A.~D.~Linde,
  Phys.\ Lett.\ B {\bf 249}, 18 (1990);
\bibitem{Adams:1990ds} 
  F.~C.~Adams and K.~Freese,
  Phys.\ Rev.\ D {\bf 43}, 353 (1991)
  [hep-ph/0504135].
  
  
\bibitem{Linde:1993cn} 
  A.~D.~Linde,
  Phys.\ Rev.\ D {\bf 49}, 748 (1994)
  [astro-ph/9307002].


\bibitem{Copeland:1994vg} 
  E.~J.~Copeland, A.~R.~Liddle, D.~H.~Lyth, E.~D.~Stewart and D.~Wands,
  Phys.\ Rev.\ D {\bf 49}, 6410 (1994)
  [astro-ph/9401011];
\bibitem{Cortes:2009ej} 
  M.~Cortes and A.~R.~Liddle,
  Phys.\ Rev.\ D {\bf 80}, 083524 (2009)
  [arXiv:0905.0289 [astro-ph.CO]].
  
\bibitem{Clesse:2014fwa} 
  S.~Clesse and J.~Rekier,
  Phys.\ Rev.\ D {\bf 90}, no. 8, 083527 (2014)
  [arXiv:1407.1984 [astro-ph.CO]].
  
\bibitem{Linde:1981zj} 
  A.~D.~Linde,
  Nucl.\ Phys.\ B {\bf 216}, 421 (1983)
  [Erratum-ibid.\ B {\bf 223}, 544 (1983)].

  
\bibitem{Linde:2005ht} 
  A.~D.~Linde,
  Contemp.\ Concepts Phys.\  {\bf 5}, 1 (1990)
  [hep-th/0503203].
  
\bibitem{Linde:1995rv} 
  A.~D.~Linde and A.~Mezhlumian,
  Phys.\ Rev.\ D {\bf 52}, 6789 (1995)
  [astro-ph/9506017].
\bibitem{Li:2003tv} 
  F.~Y.~Li, M.~X.~Tang and D.~P.~Shi,
  Phys.\ Rev.\ D {\bf 67}, 104008 (2003)
  [gr-qc/0306092].
  
  M.~l.~Tong, Y.~Zhang and F.~Y.~Li,
  Phys.\ Rev.\ D {\bf 78}, 024041 (2008)
  [arXiv:0807.0885 [gr-qc]].
  
\bibitem{Dufaux:2007pt} 
  J.~F.~Dufaux, A.~Bergman, G.~N.~Felder, L.~Kofman and J.~P.~Uzan,
  Phys.\ Rev.\ D {\bf 76}, 123517 (2007)
  [arXiv:0707.0875 [astro-ph]].
  
  
  
\bibitem{Ashoorioon:2013oha} 
  A.~Ashoorioon, B.~Fung, R.~B.~Mann, M.~Oltean and M.~M.~Sheikh-Jabbari,
  JCAP {\bf 1403}, 020 (2014)
  [arXiv:1312.2284 [hep-th]].
  
\bibitem{Ade:2013uln} 
  P.~A.~R.~Ade {\it et al.}  [Planck Collaboration],
  Astron.\ Astrophys.\  {\bf 571}, A22 (2014)
  [arXiv:1303.5082 [astro-ph.CO]].
  
\bibitem{Planck:2015xua} 
  P.~A.~R.~Ade {\it et al.}  [Planck Collaboration],
  arXiv:1502.01589 [astro-ph.CO].
  
\bibitem{Ashoorioon:2011ki} 
  A.~Ashoorioon and M.~M.~Sheikh-Jabbari,
  JCAP {\bf 1106}, 014 (2011)
  [arXiv:1101.0048 [hep-th]];
  A.~Ashoorioon and M.~M.~Sheikh-Jabbari,
  Phys.\ Lett.\ B {\bf 739}, 391 (2014)
  [arXiv:1405.1685 [hep-th]];
  A.~Ashoorioon, H.~Firouzjahi and M.~M.~Sheikh-Jabbari,
  JCAP {\bf 0906}, 018 (2009)
  [arXiv:0903.1481 [hep-th]].
  
\bibitem{Starobinsky:1980te} 
  A.~A.~Starobinsky,
  Phys.\ Lett.\ B {\bf 91}, 99 (1980).
  
\bibitem{Callan:1977pt} 
  C.~G.~Callan, Jr. and S.~R.~Coleman,
  Phys.\ Rev.\ D {\bf 16}, 1762 (1977);
  C.~G.~Callan, Jr. and S.~R.~Coleman,
  Phys.\ Rev.\ D {\bf 16} (1977) 1762;
  S.~R.~Coleman and F.~De Luccia,
  Phys.\ Rev.\ D {\bf 21}, 3305 (1980).
  
\bibitem{Adams:1993zs} 
  F.~C.~Adams,
  Phys.\ Rev.\ D {\bf 48}, 2800 (1993)
  [hep-ph/9302321].


\bibitem{Ade:2014xna} 
  P.~A.~R.~Ade {\it et al.}  [BICEP2 Collaboration],
  Phys.\ Rev.\ Lett.\  {\bf 112}, 241101 (2014)
  [arXiv:1403.3985 [astro-ph.CO]].
  

  
\bibitem{Baumann:2008aq} 
  D.~Baumann {\it et al.}  [CMBPol Study Team Collaboration],
  AIP Conf.\ Proc.\  {\bf 1141}, 10 (2009)
  [arXiv:0811.3919 [astro-ph]].
    
 
  
\bibitem{Kosowsky:1992vn} 
  A.~Kosowsky and M.~S.~Turner,
  Phys.\ Rev.\ D {\bf 47}, 4372 (1993)
  [astro-ph/9211004].

\bibitem{Huber:2007vva} 
  S.~J.~Huber and T.~Konstandin,
  JCAP {\bf 0805}, 017 (2008)
  [arXiv:0709.2091 [hep-ph]].
  
\bibitem{Ashoorioon:2004wd} 
  A.~Ashoorioon and R.~B.~Mann,
  Nucl.\ Phys.\ B {\bf 716}, 261 (2005)
  [gr-qc/0411056].
  
\bibitem{Cruise:2006zt} 
  A.~M.~Cruise and R.~M.~J.~Ingley,
  Class.\ Quant.\ Grav.\  {\bf 23}, 6185 (2006).

\bibitem{Kawamura:2011zz} 
  S.~Kawamura, M.~Ando, N.~Seto, S.~Sato, T.~Nakamura, K.~Tsubono, N.~Kanda and T.~Tanaka {\it et al.},
  Class.\ Quant.\ Grav.\  {\bf 28}, 094011 (2011).
  
\bibitem{Crowder:2005nr} 
  J.~Crowder and N.~J.~Cornish,
  Phys.\ Rev.\ D {\bf 72}, 083005 (2005)
  [gr-qc/0506015].
  
\bibitem{Lopez:2013mqa} 
  A.~Lopez and K.~Freese,
  arXiv:1305.5855 [astro-ph.HE].
  



\bibitem{Hawking:1981fz} 
  S.~W.~Hawking and I.~G.~Moss,
  Phys.\ Lett.\ B {\bf 110}, 35 (1982).
  
\bibitem{La:1989za} 
  D.~La and P.~J.~Steinhardt,
  Phys.\ Rev.\ Lett.\  {\bf 62}, 376 (1989)
  [Erratum-ibid.\  {\bf 62}, 1066 (1989)].
  
\bibitem{Easther:2006gt} 
  R.~Easther and E.~A.~Lim,
  JCAP {\bf 0604}, 010 (2006)
  [astro-ph/0601617].
  
  
\bibitem{Easther:2006vd} 
  R.~Easther, J.~T.~Giblin, Jr. and E.~A.~Lim,
  Phys.\ Rev.\ Lett.\  {\bf 99}, 221301 (2007)
  [astro-ph/0612294].
  

  
\bibitem{Ashoorioon:2013eia} 
  A.~Ashoorioon, K.~Dimopoulos, M.~M.~Sheikh-Jabbari and G.~Shiu,
  JCAP {\bf 1402}, 025 (2014)
  [arXiv:1306.4914 [hep-th]].
  
 
  



\end{thebibliography}
\end{document}